# Progressive Neural Index Search for Database System

Sai Wu
wusai@zju.edu.cn
Zhejiang University

Xinyi Yu
yuxinyi@zju.edu.cn
Zhejiang University

Xiaojie Feng
xiaojie.fxj@alibaba-inc.com
Alibaba Cloud

Feifei Li
lifeifei@alibaba-inc.com
Alibaba Cloud

Wei Cao
mingsong.cw@alibaba-inc.com
Alibaba Cloud

Gang Chen
cg@zju.edu.cn
Zhejiang University

## ABSTRACT

As a key ingredient of the DBMS, index plays an important role in the query optimization and processing. However, it is a non-trivial task to apply existing indexes or design new indexes for new applications, where both data distribution and query distribution are unknown. To address the issue, we propose a new indexing approach, NIS (Neural Index Search), which searches for the optimal index parameters and structures using a neural network. In particular, NIS is capable for building a tree-like index automatically for an arbitrary column that can be sorted/partitioned using a customized function. The contributions of NIS are twofold. First, NIS constructs a tree-like index in a layer-by-layer way via formalizing the index structure as abstract ordered and unordered blocks. Ordered blocks are implemented using $B^+$-tree nodes or skip lists, while unordered blocks adopt hash functions with different configurations. Second, all parameters of the building blocks (e.g., fanout of $B^+$-tree node, bucket number of hash function and etc.) are tuned by NIS automatically. We achieve the two goals for a given workload and dataset with one RNN-powered reinforcement learning model. Experiments show that the auto-tuned index built by NIS can achieve a better performance than the state-of-the-art index.

## 1 INTRODUCTION

Millions of users deploy various applications on the Alibaba Cloud and employ our database PolarDB[2] as their data management system. One of our crucial tasks is to optimize users' data access with a limited cloud resource, where we find that indexes play an important role. However, it is very challenging to fine-tune the index performance, currently requiring several weeks of efforts from our experienced DBA. The main causes are data diversity and variety of user access patterns. E.g., Taobao, an online shopping platform, shows very different data distributions and access patterns from DingTalk, a mobile collaboration tool. The same index configuration from Taobao does not necessarily provide a satisfied performance for DingTalk. In the ideal case, our DBA should provide a customized index configuration for every application, which is not possible for a Cloud service provider.

Fortunately, we observe that an application normally has a fixed access pattern (most queries follow some pre-defined templates and only a few ad-hoc queries) and its data also show a stable distribution. In practice, our DBAs start with an index configuration based on their experiences and gradually improve it based on the access pattern and data distribution. But they still face two challenges. First, given so many existing index structures, it is unknown which one performs best. A safe guess is $B^+$-Tree for columns requiring range search and hash index for columns requiring fast lookup. But in most cases, they are not the optimal solution. Sometimes, no existing index structure is capable of handling the unique access pattern efficiently for a specific application. We may have to design a new one. Second, even we limit our scope to popular indexes like $B^+$-Tree, skip list and hash. There are many tunable parameters, such as node size and fanout of $B^+$-Tree, the growing-up probability of skip list and the bucket number of hash function. To find a proper configuration for those parameters, we need to run a series of A/B tests, lasting for a few days.

In this paper, we propose NIS, a Neural Index Search approach, to automatically assemble an index for a given dataset and query workload, which can free our DBAs from the heavy index building and tuning work. The only assumption of NIS is that users can provide a function to sort or partition the data, which is valid for most applications. NIS formalizes various index structures into two abstract index building blocks, ordered block and unordered block. Ordered block, where keys are sorted in ascending order, can be implemented as $B^+$-Tree node or skip list. Unordered block, where keys are partitioned using customized functions, can be implemented as hash bucket. Both abstract blocks follow the format of $[key, value]^+$, where $key$ denotes the indexed key, and $value$ refers to the pointer to the next index block or the memory/disk address for the data values.

To address the two challenges (index selection/construction and index tuning), we apply the policy gradient[14] strategy to train a reinforcement learning model using RNN (Recurrent Neural Network) as backbones, which can

- Construct a tree-like index in a layer-by-layer way, where each layer is a sequence of abstract index blocks partitioning the search space with a pre-defined function.
- Search the optimal configuration for each index block, including block type, block size, minimal and maximal number of keys in a block and etc.

The first step predicts the general structure, while the second step materializes the index. In our implementations, the two steps are interleaved by stacking RNN together. Predictions of previous layers are used as context input to the RNN for the next layer, which decides whether to create a new layer or not and if a new layer is being constructed, predicts all the tunable parameters and types for each index block (ordered/unordered). In theory, NIS can produce many different tree-based, list-based and hash-based indexes via different configurations (shown in Section 2).

In this paper, we focus on the in-memory version of NIS. Compared to the disk version, in-memory NIS is more challenging because 1) the in-memory index can have multiple layers, while the disk-based one is limited to 2-3 layers; 2) since fragmentation does not apply to memory-optimized indexes, each index block can have a customized size; and 3) in-memory index is more sensitive to the access patterns. In our experiments, NIS outperforms many existing state-of-the-art in-memory indexes on various workloads. Moreover, we also provide an incremental learning mechanism for NIS. So it can handle the case where query/data distribution changes gradually over time. Our experiments show that if newly inserted data follow the same distribution, indexes generated by NIS can provide a good performance without any adjustment. If only a few portion (e.g., 10%) of new data show a different distribution, NIS employs an economic incremental learning model to adjust the index configurations. Experiments show that it still provides a better performance than others.

The idea of NIS is analogy to the NAS(Neural Architecture Search)[27]. In NAS, the hyper parameters of a neural model are tuned by another neural network. Therefore, we do not need to design a new neural model for a particular image processing task (image classification, image segmentation and object recognition) on new datasets. The parameters of CNN (convolutional neural network), e.g., kernel size, the stride size and channel number, and how the CNNs, pooling layers and normalization layers are stacked together are searched and learned automatically[12][23][11]. The auto-generated models show comparable performance to the fine-tuned models by human experts.

The closest work to ours is the *Learned Index Structures* proposed by Google[7]. It tries to learn an ordered neural mapping function for each key and stack those functions as a tree index. One challenge is to improve the prediction performance of neural models (from milliseconds to nanoseconds), invoking many engineering efforts. NIS adopts a different strategy by searching for the solution of how to combine existing index blocks and tune their parameters for specific applications. NIS does not suffer from the slow prediction of neural models, since once the index has been materialized, it can work independently.

In summary, we make the following contributions in NIS.
- We propose a neural index search framework, which applies a reinforcement learning model to search and tune new indexes for a given dataset and query workload automatically.
- We propose a conditional RNN model to generate multi-layer tree-like indexes, where each layer is an ordered list of index blocks and the construction of a layer depends on all existing layers.
- An incremental learning approach is adopted to support gradually updated query patterns.
- Experiments with state-of-the-art indexes show that the NIS generated index can achieve better performances.

The remaining of the paper is organized as follows. In Section 2, we formulate our problem and give an overview of our framework. In Section 3, we introduce our conditional RNN model and training process. We discuss our index construction and implementation details in Section 4. The NIS is evaluated in Section 5 and we briefly review some related work in Section 6. The paper is concluded in Section 7.

## 2 BACKGROUND AND OVERVIEW
### 2.1 Problem Formulations

Millions of applications are deployed on the Alibaba Cloud and many of them adopt the PolarDB as their data management systems. Once we make an agreement with our customers for the performance of their applications (e.g., average response time less than 10ms and throughput greater than 100,000/s), it is our job to tune and optimize their database instances on PolarDB.

Index tuning is one of the most important and challenging tasks that we ever met, which including two steps: identifying candidate columns for indexing and constructing a proper index for each candidate. Given a workload $W$ consisting of a set of frequently used queries, the first step picks the columns that are used as predicates for indexing. We have another work discussing the technical details. In this paper, we focus on the second step, formalized as:

**Problem Definition**: Given a column $C$ and workload $W$, assume that $C$ can be sorted by a function $f$. How can we generate a proper tree-like hierarchical index structure $I$ for $C$, which is tuned to minimize the total processing latency

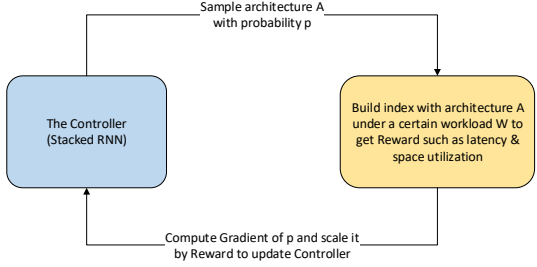

Figure 1: General Idea of NIS

of $W$ with a storage budget $B$, computed as:
$$c_t = \sum_{\forall q_i \in W} f(q_i) \times w(q_i)$$

$f(q_i)$ returns the latency of processing $q_i$ with the existence of $I$. $w(q_i)$ represents the weight of $q_i$, which is set as the frequency of $q_i$ in $W$ currently. The total storage cost of the index should be less than $B$.

We do not intent to invent new index structures. Instead, our plan is to reuse existing index structures and fine-tune them for our target workloads. Therefore, we introduce two abstract index blocks that can be materialized as popular existing index structures.

*Definition 2.1.* The ordered index block is described as $I_o = \{S, F, [L, U), pt\}$, where $S$ denotes a sorted list of keys, $F$ is the mapping function, $[L, U)$ denotes the key range of the block and $pt$ points to the next sibling block. For $I_o$, we have the properties:
- $\forall S[i] \in S, L \leq S[i] < U$.
- $S[i] \leq S[j]$, if $i < j$.
- $F(S[i]) \leq F(S[j])$ if $i < j$

In the ordered index block, the mapping function $F$ is maintained as a list of sorted key-value pairs $[(k_i, v_i)^*]$, where $k_i < k_j$ for $i < j$ and $v_i$ refers to the position of the index block in the next layer or the real data values.

*Definition 2.2.* The unordered index block is denoted as $I_u = \{S, H, [L, U), pt\}$, where $S$ is a set of keys, $H$ is a hash function(currently, we use the standard SHA-1 hash function), $[L, U)$ denotes the key range of the block and $pt$ points to the next sibling block. For $I_u$, we have $\forall S[i] \in S, L \leq S[i] < U$.

Each index block can hold up to $m$ keys. $m$ can be calculated by the key size and the cache line size(for memory index) or block size(for disk index). However, the initial number of keys inside each block(denoted as $x$) is a tunable parameter, which is learned by the NIS through training. Table 1 lists the hyper-parameters learned by the NIS. In unordered index block, the whole block is maintained as a hash table, where we have maximal $m$ buckets and the buckets maintain pointers to index blocks in next layer.

Table 1: Tunable Hyper-Parameters

| Block Type | ordered/unordered |
|---|---|
| $x$ | the initial number of keys in an index block |
| $y$ | the number of blocks in a group |
| $\alpha$ | the block will split when more than $\alpha m$ keys |
| $\beta$ | two blocks will merge when both less than $\beta m$ keys |
| $\gamma$ | a probability vector for creating skip links |

## 2.2 Overview of NIS

The design of NIS follows the same philosophy of the NAS. Figure 1 shows the general architecture. Given a database $D$ and query workload $W$, NIS employs a controller to tune the hyper-parameters listed in Table 1 and decides how the abstract index blocks can be assembled as an index. In this paper, the controller is a reinforcement-learning model, which applies the policy gradient[14] to update status and is implemented as stacked RNN. After the controller makes a prediction, an index builder materializes the corresponding index and deploys it on the database. We test the query workload $W$ using the index to get the latency and space utilization as our rewards, which are used as feedbacks for the controller to update its predictions. The process continues, until the latency and space utilization converge.

One challenging of applying NIS to predict the index structure is the scalability. Suppose we have 10 million keys and each block can hold up to 1000 keys. We need at least 10,000 blocks to maintain those keys. In other words, the NIS needs to generate hyper-parameters for a large number of blocks sequentially. However, existing neural models are not capable of predicting such a long sequence. To reduce the prediction cost of NIS, we classify the blocks into groups.

*Definition 2.3.* An index block group $G$ is a set of index blocks responsible for consecutive key ranges and sharing the same hyper-parameters.

Given a set of keys in $[min, max)$, to generate a new index, the controller first creates a layer with one index block. The type of the block and the initial size of its keys are all decided by the controller. Suppose the block partitions the key ranges into $P_0 = [k_0, k_1), P_1 = [k_1, k_2),...,P_{n-1} = [k_{n-2}, k_{n-1})$(namely, $x$ is set as $n$ by the controller). By default, we partition the key range evenly. The controller starts building the second layer of index for each range by adaptively generating an index block group. The hyper-parameters of a group is learned by the controller. In other words, $y$ index blocks are created for each group and the corresponding key range is partitioned into $y$ non-overlapped consecutive ranges, one for each block. Inside each group, to facilitate the query, each index block can create skip links up to $\log y$ blocks inside the same group. In particular, suppose current group $G$ has $y$ index blocks: $\{I_0, ..., I_{y-1}\}$. For block $I_a$, it will create a skip link to $I_b$ with a probability $p_i$, if $b = a + 2^i (0 < i \le \log y)$ and $b < y$. The probability $p_s$ is estimated by the controller, and we have $\gamma = \{p_1, p_2, ..., p_{\log y}\}$.

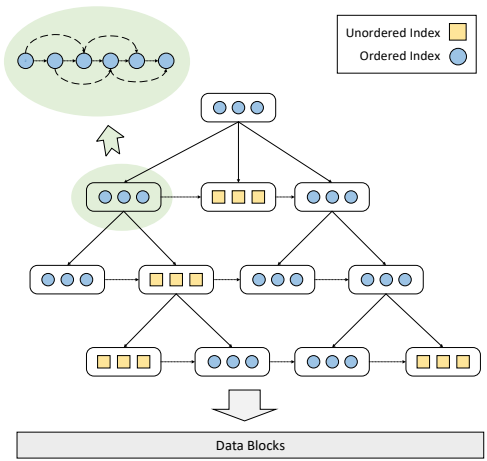

Figure 2: An Example Index Generated by NIS

The above index construction process continues for each index block, until no key range has more than $\beta m$ keys. In this way, we may generate an imbalanced index search tree with skip links inside each block group.

The controller outputs its final decision as a sequence of operations, which are read by the index builder for construction. The index builder adopts a streaming approach to materialize the index. In particular, it first creates an abstract index by stacking the index blocks predicted by the controller. Then, it reads in the data and applies the data stream to materialize the index. If block $b_i$ is split into $b_0$ and $b_1$, both $b_0$ and $b_1$ will share the same hyper-parameters of $b_i$. However, if the controller generates a good prediction, we do not need to split data blocks frequently. To speed up the index construction, the index builder employs multiple threads to assemble the index. Finally, the index is deployed on the database and tested against the given workload.

As an example, Figure 2 shows an index generated by NIS, which performs better than an ordinary B$^+$-tree for queries following Zipfian distribution. The blue and yellow nodes represent the ordered and unordered index blocks, respectively. The dashed line denotes the skip links inside a group. To process a query, we start from the root block as searching a B$^+$-tree. When reaching a group, we pick the skip links to simulate the search process as the skip-list.

By learning different hyper-parameters, NIS can simulate different types of conventional indexes, such as:

**B$^+$-tree** All the index blocks are ordered blocks with the same configuration of $m$ and $\alpha$ and $\beta$ are set to 1 and 0.5 respectively. $y$ is set as 1 for all groups.

**Hash** All the index blocks are unordered blocks and the index only has one layer.

**Skip-list** Each layer only has one index group and the upper layer group has fewer blocks (a smaller $y$).

It can be seen that the large search space of NIS allows us to explore more new index structures by combining different existing index structures for specific workloads and datasets.

## 3 IMPLEMENTATION OF CONTROLLER

In this section, we show how the controller learns to predict the hyper-parameters for the index. We first discuss the architecture of our neural model and then elaborate on how the training process works.

### 3.1 The Architecture of Controller

We consider the prediction of hyper-parameters as a task of sequence prediction. Therefore, the backbone of our controller is an RNN network powered by the LSTM[1]. Figure 3 illustrates the basic architecture of the controller. The controller consists of multiple layers of LSTMs to predict a tree-like index. The new layer will use the hidden states of previous layers as the context during its prediction.

Our model consists of a basic building block as shown in Figure 3. The neural block predicts the six hyper-parameters (*Block Type*, $x$, $y$, $\alpha$, $\beta$ and $\gamma$) for an index block group as a sequence via the RNN model. The whole block consists of three neural layers, an embedding layer, a LSTM layer and a softmax layer.

The bottom layer is an embedding layer, formalizing the input as a binary vector representation. For the first state of the RNN, the min/max values of the keys, the number of unique keys and a coarse histogram are transformed into binary vectors and concatenated together as the input. For the following states, the generated vector from the softmax layer for previous state is used as the input.

The middle layer applies the LSTM to learn the correlations between different states. Selections of previous hyper-parameters affect the choices for the following ones. In fact, we also tested the biLSTM(bi-directional LSTM), but did not find a significant improvement. So we stick to the basic LSTM model.

The top layer is a softmax layer for prediction. We transform our task into a classification problem by creating a set of pre-defined values for each hyper-parameter and only allowing the neural model to pick one of the pre-defined values. Table 2 lists our current pre-defined values. The parameter $\gamma$ has no default values, since it depends on the value of $y$. Specifically, we will generate a probability vector $\{p_1, p_2, ..., p_{\log y}\}$ indicating whether to create skip links to the neighboring $\log y$ blocks in the same group. The $i$th skip link will be established with a probability $p_i$.

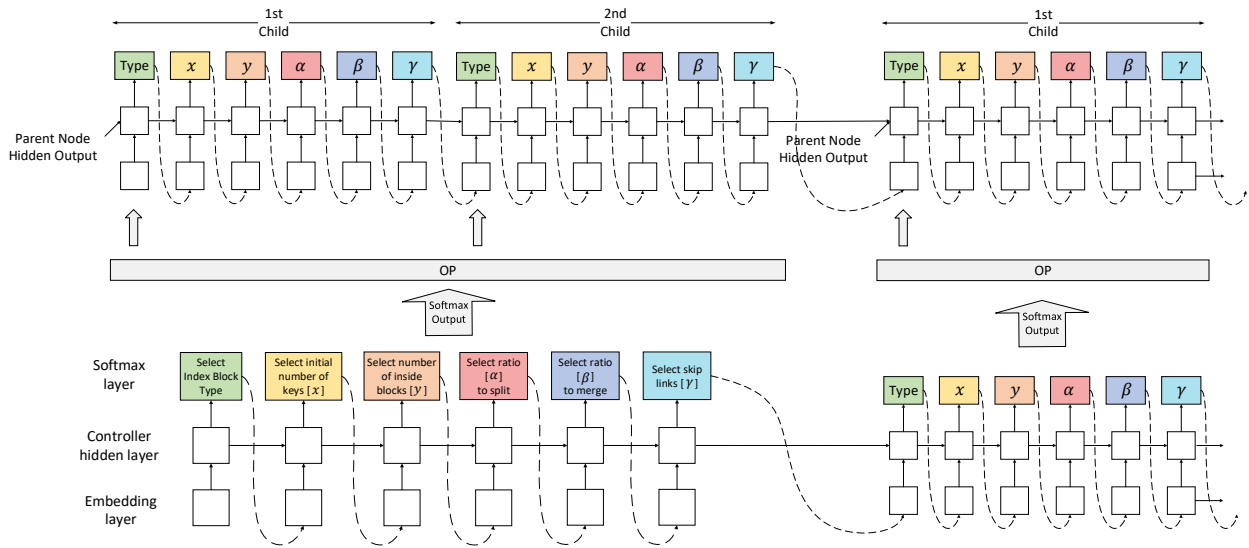

Figure 3: The Neural Model of NIS

Table 2: Pre-defined Values

| Hyper-Parameters | Values |
|---|---|
| $x$ | $\frac{m}{4}, \frac{m}{2}, \frac{3m}{4}, m$ |
| $y$ | 32, 64, 128, 256 |
| $\alpha$ | 0.5, 0.6, 0,7 0.8, 0.9, 1 |
| $\beta$ | 0.1, 0.2, 0.3, 0.4 |

The basic neural blocks are chained together to predict the hyper-parameters for index block groups. After successfully generating the prediction for one layer of index, the controller can start up a new layer, if some index blocks need to be further partitioned. Then, the hidden states of current layer are used as the context during the prediction for the next layer.

Figure 3 illustrates the idea. The outputs from the softmax layer of a neural block are concatenated together and used as input to the neural blocks in the next layer. The estimated number of children of a neural block is $x$, and hence, its hidden states will be used as contexts for $x$ consecutive neural blocks in the following layer. In this way, we can progressively generate the hyper-parameters for a multi-level index, which, in fact, can simulate most existing index structures.

### 3.2 Training the Controller

The hyper-parameters generated by the controller can be considered as a series of operators $a_{1:t}$ (operators from the start to time $t$) which are used to construct a new index for a given workload $W$ and database $D$. At convergence, the new index is integrated into $D$ and we test it against the workload $W$. The total processing time $c_t$ and the index space utilization cost $c_s$ (the percentage of the index that has been used for maintaining keys) are our main concerns. So the reward is defined as:

$$R = \rho \frac{c_b - c_t}{c_b} + (1 - \rho) c_s$$

$c_b$ is the baseline processing time without any index and $\rho$ is a tunable parameter to balance the importance of the two terms. We have conducted experiments to show the effect of $\rho$.

To find the optimal index, we ask the controller to maximize its expected reward, represented as $J(\theta_c)$. We use $\theta_c$ to denote all parameters of the controller model. We have:

$$J(\theta_c) = E_{1:T;\theta_c}[R]$$

The reward $R$ is achieved by monitoring the performance of database $D$ and hence, is not differentiable. We apply the policy gradient strategy as in [14]:

$$\nabla J(\theta_c) = E_{P(a_{1:T};\theta_c)} \sum_{t=1}^{T} (R-b)\nabla \log P(a_t|a_{t-1}) \quad (1)$$

$T$ is the total number of hyper-parameters for predicting and $b$ is a baseline function to reduce the variance of the reward. Currently, let $\mu$ be the aging factor. $b^{(n)}$ in $n$th training sample is recursively defined as

$$b^{(n)} = \begin{cases} 0 & \text{if } n = 0 \\ \mu b^{(n-1)} + (1-\mu)R^{(n)} & n \geq 1 \end{cases}$$

As a result, $b^{(n)}$ can be estimated as:

$$b^{(n)} = (1-\mu)(R^{(n)} + \mu R^{(n-1)} + \mu^2 R^{(n-2)} + ... + \mu^{t-1}R^{(1)})$$

Equation 1 shows how the parameters $\theta_c$ of the controller network are updated based on the reward $R$ gradually, which is represented as

$$\theta_c := \theta_c + \sigma \nabla J(\theta_c)$$

where $\sigma$ is the learning rate.

In practice, we use the exhaustive weighted summation form to replace the expected value in Equation 1. If we have enough training samples, we can estimate the $\nabla J(\theta_c)$ as ($N$ is the batch size of the controller):

$$\nabla J(\theta_c) = \frac{1}{N} \sum_{n=1}^{N} \sum_{t=1}^{T} (R^{(n)} - b^{(n)}) \nabla \log P(a_t^{(n)}|a_{t-1}^{(n)}) \quad (2)$$

In our experiments, we find that a small $N$ is good enough for the model to converge to a satisfied result.

The intuition of policy gradient is to increase the probability of $P(a_t|a_{t-1})$, if $R-b$ is positive. Otherwise, we decrease the probability. However, during the training process, we find that if $P(a_t|a_{t-1})$ is large enough, $R-b$ will be always positive (because the model gives up on exploring new results and sticks to current sub-optimal one), causing $P(a_t|a_{t-1})$ to converge to 1. On the contrary, if $P(a_t|a_{t-1})$ is very small, $R-b$ will be negative in most estimations, and hence, $P(a_t|a_{t-1})$ will converge to 0. In both cases, we obtain a local optimal results. To avoid such problems, we clip the sample data and only update the probabilities within $[\epsilon, 1-\epsilon]$, where $0 < \epsilon \ll 1$.

Similar to other policy gradient approaches, the training process lasts for days, since we need to build each predicted index and performs benchmarking to gather corresponding rewards. To speed up the training process, we apply two optimization techniques.

We generate a set of probabilities after the softmax layer of the controller. They are used to select the hyper-parameters. For example, we obtain the probabilities $[p_1, p_2, p_3, p_4]$ for parameter $y$, indicating that we may set $y$ as 32, 64, 128, 256 with probabilities $p_1, p_2, p_3$ and $p_4$, respectively. In value-based learning approach, it was shown that random exploration can speed up the convergence. We adopt this approach for our policy-based approach. In particular, we ask the controller to ignore the generated probabilities and randomly pick a value for a hyper-parameter with a pre-defined probability $\lambda$. Initially, $\lambda = 1$ to allow a fast random exploration and gradually, we decrease $\lambda$ to 0.

The training of vanilla policy gradient approach is extremely slow due to a large exploration space. A new approach, PPO(Proximal Policy Optimization) [17], can be used to facilitate the parameter updates. In policy gradient, we update model parameters, only when we obtain new training samples. This strategy is called "on-policy" strategy. Instead, in PPO, we create a new controller model $\theta_c'$, which is employed with the environment(in our case, the index builder and database) to get training samples. The training samples obtained from $\theta_c'$ are repeatedly used by the real model $\theta_c$, so that parameters of $\theta_c$ get multiple updates for one sample. This is called "off-policy" strategy.

Using PPO, Equation 1 is rewritten as:

$$E_{P(a_{1:T};\theta_c')} \sum_{t=1}^{T} \left[ \frac{p_{\theta_c}(a_t|a_{t-1})}{p_{\theta_c'}(a_t|a_{t-1})} (R^{\theta_c'} - b) \nabla \log p_{\theta_c}(a_t|a_{t-1}) \right] \quad (3)$$

However, if the distribution of $\theta_c'$ and $\theta_c$ differs a lot, the approach may not work. So, PPO introduces the KL-divergence to balance the difference between two distributions. The equation is further revised as:

$$\begin{aligned} J_{PPO}^{\theta_c'}(\theta_c) &= J^{\theta_c'}(\theta_c) - \phi KL(\theta_c, \theta_c') \quad (4) \\ &= E_{P(a_{1:T};\theta_c')} \sum_{t=1}^{T} \left[ \frac{p_{\theta_c}(a_t|a_{t-1})}{p_{\theta_c'}(a_t|a_{t-1})} (R^{\theta_c'} - b) \right] \\ &\quad - \phi KL(\theta_c, \theta_c') \end{aligned}$$

### 3.3 Incremental Updating

As mentioned before, most applications on PolarDB have a stable data and query distribution. So we can learn an index to achieve a good performance. However, data and query will slowly evolve. Hence, we design an incremental updating model, which is a by-product of the controller. In other words, we reuse training samples from controller to build the incremental updating model.

The intuition of incremental updating model is to learn a performance prediction function $G$. Given a data distribution $D$, query distribution $Q$ and specific index $I$, $G(D, Q, I)$ returns the estimated processing latency of $Q$. We use equiwidth histograms to maintain data distribution and query distribution. We generate a vector representation for each index block by encoding its configuration parameters. Finally, $G$ is learned through a tree structured LSTM model [18]. Figure 4 shows the architecture of our model.

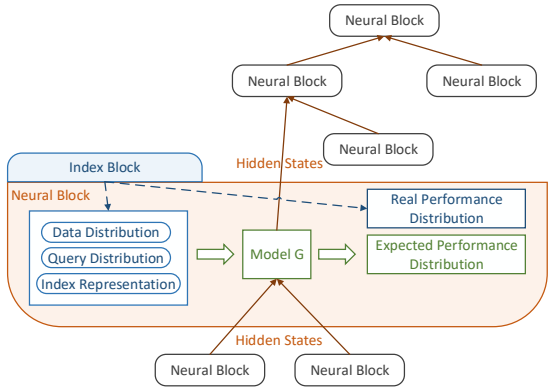

Figure 4: A Tree-LSTM for Incremental Updating

We train a neural block for each index block and connect them as a tree structure, where parent node accepts hidden states from its child nodes as context for prediction. All neural blocks actually share the same hyperparameters trained with samples from controller. During the training of a controller, we obtain an index structure under a specific data distribution $\mathbb{D}$ and query distribution $\mathbb{Q}$. We collect the statistics of the processing latency for each index block. If a block is a leaf block, its latency is its total cost of processing queries. If the block is an internal block, its latency includes both its own cost and the accumulative costs of all its descendants. Then, we train a performance prediction network to predict the cost of each index block. The network accepts $\mathbb{D}$, $\mathbb{Q}$, vector representation of the block and hidden states from child nodes as input. It outputs a hidden state which is further applied to generate predictions for internal nodes. Since we do not need a very precise estimation of latency, we transform the prediction into a classification problem, where we create 100 different performance classes and pick the class with maximal probability.

We consider queries in $\mathbb{Q}$ as a training batch and also propose a batch loss function. The loss function is based on the KL-divergence:

$$loss = \sum_i P(i) \log \frac{P(i)}{Q(i)}$$

where $i$ is a performance class, $P(i)$ and $Q(i)$ indicate how many queries are assigned to class $i$ in the prediction and real statistics.

The tree-LSTM is trained together with the controller. Then, it is applied to help us identify performance outliers. During the query processing, we collect the statistics of our index and use it to make a prediction for the performance periodically. Let $x_0$ be the initial performance of an index block after the last update. We use $x_t$ and $x'_t$ to denote its predicted performance and real performance at the $t$th epoch. An index block is marked as an outlier if either $x_t \neq x'_t$ for more than $\tau$ epochs, or $x_t - x_0 > \omega x_0$. $\tau$ and $\omega$ can be tuned to balance the index tuning cost and processing cost.

To reduce the tuning cost, we identify the outliers in a bottom-up way. If an index block is outlier, we continue to check its parent. If all child blocks are not outliers, we stop the check for this block. For those marked as outliers, we invoke the controller to find a new index structure, while for the rest index blocks, we keep their existing configurations. To speed up the learning, instead of processing queries against the new index to collect latencies, the controller asks function $G$ to obtain an estimated performance. During the updating process, we reserve the old outlier blocks to support queries. Only when the new ones have been learned, will we replace the old ones.

## 4 PROCESSING OF THE INDEX

In this section, we introduce how the predicted index can be materialized as a physical index, and how the index can be applied to process queries and updates.

### 4.1 Index Materialization

The index construction is performed in two steps. In the first step, the index builder loads the hyper-parameter predictions from the controller, which are indexed in a key-value store, to build a logical index. The logic index establishes the general structure of the index, but cannot support queries. In the second step, index builder scans data and feeds them into the logical index in a streaming way. The logical index fills in detailed key ranges and builds necessary pointers, which are finally materialized as a physical index.

*4.1.1 Logical Index.* To create the logical index, we first set up the parent-child relationships between index groups. The left index in Figure 5 shows a logical index (to simplify the digram, the skip links are not shown). During the construction of the logical index, we create the parent-child pointers, the key ranges of each index blocks and the skip links inside each index block group. Algorithm 1 shows the workflow of how the logical index is established. If parent node is an ordered block, we split its key range evenly and assign to each child block group (line 3-5). Otherwise, child block groups will share the same key range with their parent block, because the hash function will project keys into random blocks (line 7-9).

Algorithm 2 illustrates how an index block group is set up. First, it partitions the key range evenly and generates a fixed number of index blocks based on the predicted hyperparameters (line 3-6). Then, it creates multiple skip connections with the probabilities specified in the hyper-parameters (line 7-16).

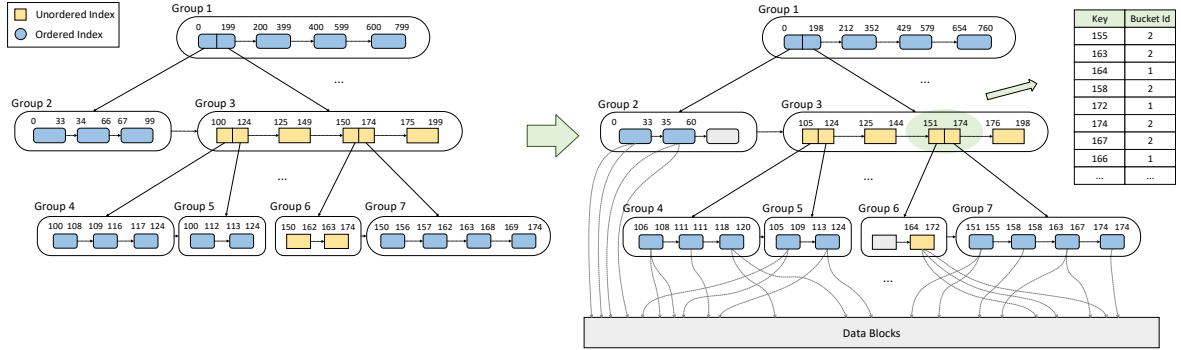

Figure 5: Materialization of the Index

**Algorithm 1** LogicalIndex(ParentBlock $p$, ParameterIndex $in$)

1: **for** $i$ = 1 to $p.x$ **do**
2:     **if** $p$ is ordered block **then**
3:         $r = [p.L + \frac{(p.U-p.L)(i-1)}{p.x}, p.L + \frac{(p.U-p.L)i}{p.x})$
4:         $g$ = CreateIndexBlockGroup($r$, $i$, $in[i]$)
5:         $p.child[i] = g$
6:     **else**
7:         $r = [p.L, p.U)$
8:         $g$ = CreateIndexBlockGroup($r$, $i$, $in[i]$)
9:         $p.child[i] = g$

**Algorithm 2** CreateIndexBlockGroup(Range $r$, Index $i$, ParameterIndex $in$)

1: $params$ = $in$.getHyperParameters($i$)
2: List $group$ = $\emptyset$
3: **for** $j$=1 to $params.y$ **do**
4:     $r' = [r.L + \frac{(r.U-r.L)(j-1)}{params.y}, r.L + \frac{(r.U-r.L)j}{params.y})$
5:     IndexBlock $B$ = new IndexBlock($params$, $r'$)
6:     $group$.add($B$)
7: **for** $j$=0 to $params.y - 1$ **do**
8:     IndexBlock $B$ = $group$.get($j$)
9:     **for** $k$=1 to $\log y$ **do**
10:         $target = j + 2^k$
11:         **if** $target > params.y$ **then**
12:             break
13:         $\bar{p}$ = roll a dice
14:         **if** $\bar{p} \geq params.\gamma[k]$ **then**
15:             IndexBlock $B'$ = $group$.get($target$)
16:             create a skip link between $B$ and $B'$

If all index blocks are ordered blocks, the generated index structure is similar to the B$^+$-tree. However, if unordered index blocks are adopted, the index structure becomes a hybrid one. For the left index in Figure 5, suppose $x$ is 2 for the root index block group. The first index block group and second index block group at level 2 are responsible for the key range [0, 99] and [100, 199], respectively. The first group is ordered blocks and suppose its $y$ is set as 3. We create three ordered index blocks inside the group and partition the key range evenly as [0, 33], [34, 66] and [67, 99]. The second index block group in level 2 is unordered index group. If $y$ = 4, we will create four unordered blocks by partitioning the key range [100, 199] uniformly. Because unordered block applies hash functions to map keys to its child block groups, the child block groups share the same key range with their parent. For example, group 4 and group 5 all have the key range [100, 124]. However, group 4 creates 3 blocks and group 5 creates 2 blocks. Note that in our index, two index block groups may have overlapped key ranges, but for the index blocks in the same group, they always maintain sorted non-intersected ranges.

*4.1.2 Physical Index.* In the second phase, the index builder loads data from the disk and feeds them to the logical index to materialize the index. The right index in Figure 5 shows the corresponding physical index for the left logic index.

The materialization process mainly handles three tasks:

(1) Update key ranges of index blocks. The key range of each block in logical index is just a rough estimation. During the materialization process, we maintain a set of [Min, Max] values for each block, indicating the actual minimal and maximal keys in each key range. After all data have been processed, we shrink the key ranges of an index block by the values. This helps us reduce the search cost by filtering the blocks as early as possible.
(2) Set up hash tables for unordered blocks. As shown in Figure 5, when data are streamed over an unordered block, we will set up the corresponding hash table. Suppose there are $x$ child index block groups and the next key is $k$. $k$ will be routed to the $i$th group, where $i = hash(k) \% x$. To help the search, we also create a

**Algorithm 3** Materialize(BlockGroup $G$, Tuple $T$, Offset $O$)

1: **if** $G$ is ordered block group **then**
2:     $B$=G.findOverlappedBlock($T.k$)
3:     **if** $B$ is at the bottom level **then**
4:         $B$.insert($T.k$, $O$)
5:     **else**
6:         $i$=$B$.find($T.k$)
7:         $B[i]$.updateMinMax($T.k$)
8:         Materialize($B[i]$, $T$, $O$)
9: **else**
10:     $B$=findOverlappedBlock($T.k$)
11:     **if** $B$ is at the bottom level **then**
12:         $B$.hash[$T.k$] = $O$
13:     **else**
14:         $i$=$B$.hash($T.k$) % $B.x$
15:         $B[i]$.updateMinMax($T.k$)
16:         $B$.updateBloomFilter($T.k$)
17:         Materialize($B[i]$, $T$, $O$)

bloom filter for each unordered block to check whether a key exists or not.

(3) Create pointers to disk data. When a key is routed to the bottom level of the index, we will create a pointer from the key to the disk offset of the corresponding record. For secondary index, one key may refer to multiple records. Then, we will merge them as a sorted list for disk offsets.

Algorithm 3 summarizes the whole materialization process. For a new tuple $T$ and its disk offset $O$, we first retrieve the block whose key range overlaps with $T.k$. If current group is at the bottom level, we just insert the key and its offset. Otherwise, we forward the tuple to the corresponding child group. The same process repeats for the unordered block group. The only difference is that we apply the hash function to map the tuple to a specific child group.

The right index in Figure 5 is the materialized index for the left one. We can find that the key ranges of index blocks are shrunk. E.g., the ranges of group 1 change from [0, 199], [200, 399],...,[600, 799] to [0,198], [212,352],...,[654,760]. The hash tables have been set up (we show the hash table of the third block in group 3). Note that we do not need to maintain the hash tables explicitly. We only need to know which hash function is being applied. Finally, we create the links from the keys to their disk offsets in the bottom level. Note that in Figure 5, the gray nodes indicate that the nodes are empty, since their key ranges do not contain any keys.

During the materialization, an index block at bottom layer may be overloaded during the materialization process, triggering the block splitting operation. We will discuss this issue in our index update section.

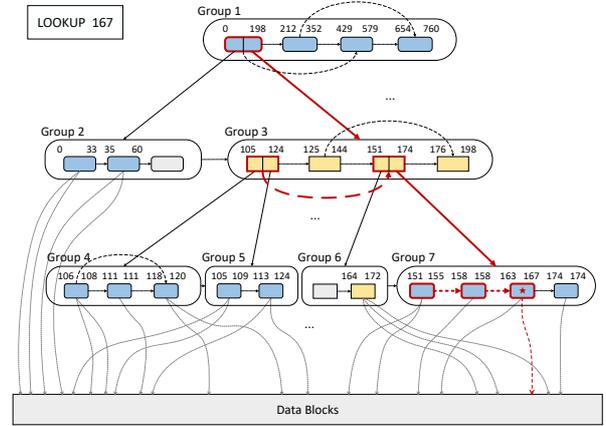

**Figure 6: Processing of Lookup Queries**

### 4.2 Search via Index

After the index has been materialized, we can apply it to process queries. In this paper, we focus on the lookup and range queries. As a hybrid index, our search process is a combination of B$^+$-tree, Hash and skip-list. We use two examples in Figure 6 and Figure 7 to demonstrate the lookup and range queries, respectively.

Suppose a lookup query retrieves the key "167". The search process works as follows. We first check the root block group(Group 1). Because the second key range of the first block contains 167, we route the query to the second child index group of the first block(Group 3). In group 3, the key ranges of the first block do not contain the key. So we route the queries based on the skip link to the third block. Then, we apply the hash function to retrieve the next index group(Group 7). Before forwarding the query to group 7, we also test it against the bloomfilter of the block. If bloomfilter returns a positive result, we continue the query in group 7. Since no skip link is set up in group 7, we scan the blocks one by one until reaching the third one, where the key is located.

Algorithm 4 gives the pseudo code for the lookup. The function *SkipListSearch* simulates the search of skip list, where we follow the skip link which points to the block satisfying: either its key range contains the key or its maximal key is the largest maximal key smaller than the search key.

For the range query [118, 124] in Figure 7, we start the same process as the lookup query. The main difference is how the query is processed when reaching an unordered block group(Group 3). The query overlaps with the first block. But a hash function may distribute the keys to all the child block groups (Group 4 and Group 5). So the range query should be forwarded to both groups. Inside each group, we follow

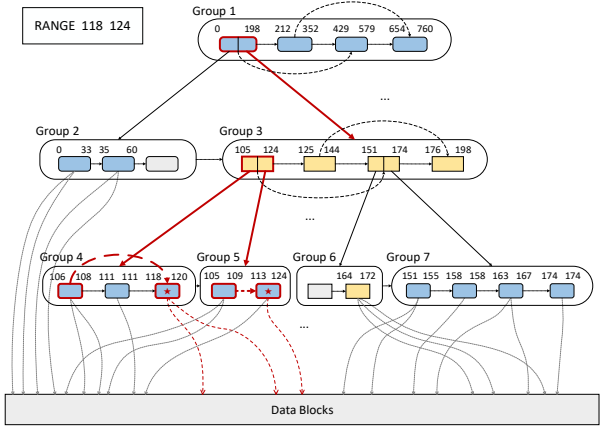

Figure 7: Processing of Range Queries

---

**Algorithm 4** Lookup(Key $k$, Group $G$)

1: $B = G$.SkipListSearch($k$)
2: **if** $B$ is the ordered block **then**
3:     **if** $B$ is at the bottom level **then**
4:         return $B$.find($k$)
5:     **else**
6:         $i=B$.find($k$)
7:         Lookup($k$, $B[i]$)
8: **else**
9:     **if** not $B$.bloomfilter($k$) **then**
10:        return NULL
11:    **if** $B$ is at the bottom level **then**
12:        return $B$.hash[$k$]
13:    **else**
14:        $i=B$.hash[$k$] % $B.x$
15:        Lookup($k$, $B[i]$)

---

the skip link to locate the smallest key and then scan the remaining blocks until reaching the largest key. We discard the details of range search algorithm.

### 4.3 Update of the Index

During the index materialization, blocks at the bottom level may be overflowed, if more than $\alpha m$ keys are inserted. This will trigger a node splitting operation. If the controller generates a good estimation for the data distribution, this problem can be partially avoided by generating an index tailored for the data distribution.

One key design philosophy of the NIS is that the index does not need to be balanced. It may create more levels of blocks for high-density data partition to facilitate the search. The unbalanced assumption reduces the complexity of node splitting and merging, since we can limit node splitting and merging to the groups at bottom level and do not propagate to the upper level.

When a bottom index block has more than $\alpha m$ keys, we split the block evenly into two new blocks. The new blocks share the same hyper-parameters as they reside in the same group. The splitting does not affect the parent block, since the key range of the index group does not change. However, we need to update the skip links, since new blocks are inserted into the group.

The merge process follows the same strategy as the split one. Two consecutive blocks are merged together when both blocks have less than $\beta m$ keys. And we update the skip links of the new block.

During the insertion and deletion, we also need to handle the changes of key ranges. As shown in Figure 5, the initial key ranges are setup during the materialization process. When a new key "150" is inserted, no existing index blocks can hold the key. So we need to find the nearest block to expand it key range. We called the process, *expanding*. The nearest block is defined as block $B$ in the group with the minimal min $(|B.L - k|, |B.U - k|)$, where $[B.L, B.U]$ is $B$'s key range.

In Figure 5, the insertion process invokes the expanding function for group 3, who finds the closest block to key "150" is the third one and expands its range from [151, 174] to [150, 174]. Since this is an unordered block, it applies the hash function to decide which child index group should handle the insertion. Suppose it is group 6, who has one empty block and one block for [164, 172]. The empty block has the highest priority during the expanding process. So key "150" will be stored at the empty block, which updates its range as [150, 150]. The deletion process follows the same strategy by introducing a range shrinking processing. We will not elaborate the details.

## 5 EXPERIMENTS

We train the NIS using two servers sharing the same hardware configurations (Xeon CPU 32 cores, 64GB DDR3 RAM, 2MB L2 and 20MB L3 and NVIDIA GTX TITAN X). One server is dedicated to the training process of controller and the other one is used for index materialization and evaluations. We employ four datasets for evaluations: a synthesis uniform dataset (*uniform64*) and three real datasets (*amzn, facebook, osmc*). All datasets have 200 million keys. Detailed descriptions of the datasets can be found in [5].

For comparison purpose, we use the open-sourced implementations of B$^+$-Tree, SkipList, ART[9], FAST[3], Bw-Tree[21] and Learned Index[7] (denoted as RMI) as our baselines[1]. All indexes are in-memory indexes and no disk I/Os are involved.

---

[1] https://github.com/learnedsystems/rmi

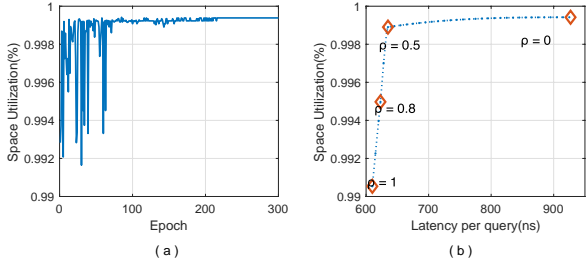

Figure 8: Training of the Controller

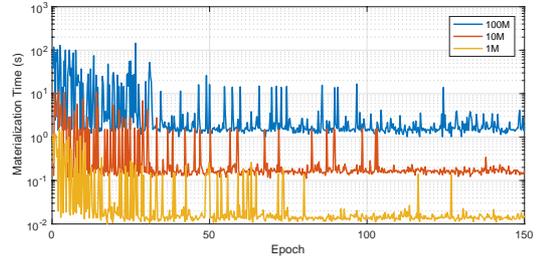

Figure 9: Cost of Index Materialization

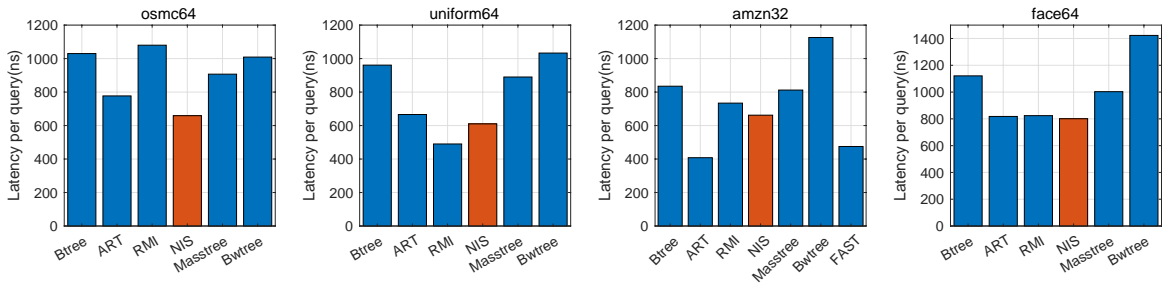

Figure 10: Performance of Lookup-Only Workload

## 5.1 Training the NIS

We first show the training performance of the controller for the *uniform64* dataset with read-only workloads. During our training, the batch size of the controller is set as 64. Namely, it will generate 64 different index structures. To obtain proper rewards, those indexes are materialized by the index builder and deployed to be evaluated against the pre-defined workloads. The training process for one batch is considered as an epoch for the learning model, typically lasting from a few minutes to less than one second (mainly for index materialization and evaluation). This is because when the index structure is close to convergence, its processing cost is significantly reduced. As shown in Figure 8(a), the controller requires about 100 epochs to converge.

The reward function used in the controller is a combination of the processing latency and index space utilization cost. We use parameter $\rho$ to tune the weights of the two terms. Figure 8(b) shows the effect of different reward functions. When $\rho = 0$, we only consider the space utilization and hence the generated index is almost full. On the other hand, if $\rho = 1$, the latency is the only concern. We observe that by tuning parameter $\rho$, we can achieve a trade-off between the latency and space utilization. In the following experiments, $\rho$ is set as 1 to minimize the search latency.

In Figure 9, we show the cost of index materialization, namely, the latency of transforming a logic index into physical index. We test 1 million to 100 million data. The materialization cost reduces during the training, since the index structure has been cached and is only partially updated in the following epochs.

## 5.2 Performance of Lookup-Only Workload

In this experiment, we compare the NIS with other baseline approaches. We generate 10 million single key lookup queries following the same distribution with the corresponding dataset. Namely, we have more queries for high-density data ranges. Figure 10 shows the performances of different approaches on different datasets. The y-axis denotes the average processing cost of queries in nano-seconds. *osmc* is the most complex dataset and hence, NIS performs much better than the other approaches, indicating that it can tune the index structure based on data and query patterns. For the other two real datasets, *amzn* and *facebook*, NIS still performs better than the RMI. However, for the uniform dataset, RMI achieves a better performance. This is because of the complex structure of NIS and the neural network may not fully converge. We also find that some state-of-the-art indexes, such as ART and FAST, are quite good at processing lookup requests (the open-sourced FAST does not support 64 bits keys, so we only show its performance on 32 bits dataset).

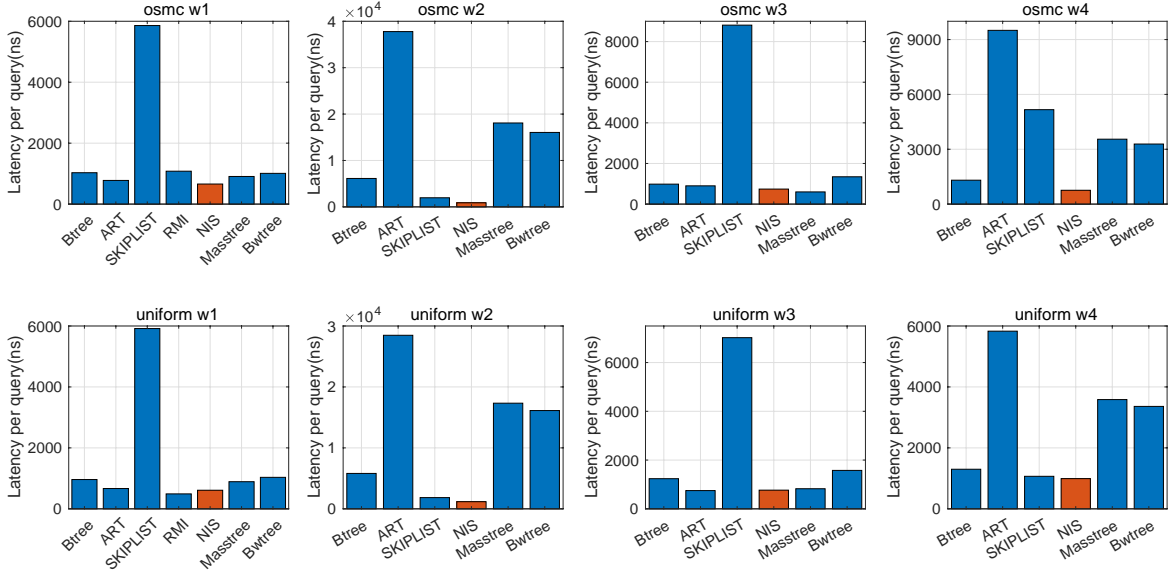

Figure 11: Performance of Mixed Workload

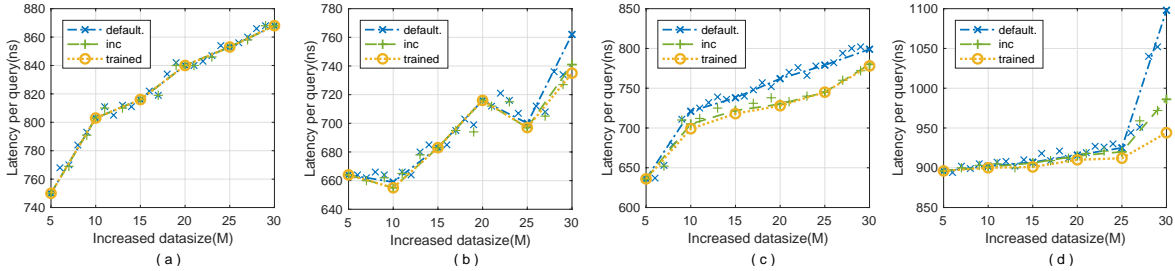

Figure 12: Incremental Learning

## 5.3 Performance of Mixed Workload

In this experiment, we focus on two datasets, *osmc* and *uniform*, and generate four different workloads. W1 denotes the workload that we used in previous experiment (10 million lookup queries). W2 contains 1 million range queries with a selectivity 1%. W3 mixes 5 millions lookups and 5 millions insertions operations. W4 mixes 2 millions lookups, 2 millions insertions and 1 million range queries (selectivity=1%). We show the average query processing cost in nano-seconds. Figure 11 shows the results. RMI is only shown in W1, because current RMI implementation does not support range queries and updates. For the mixed workload, NIS shows a superior performance than the other indexes even on the uniform dataset, indicating that it can be used to support various application scenarios.

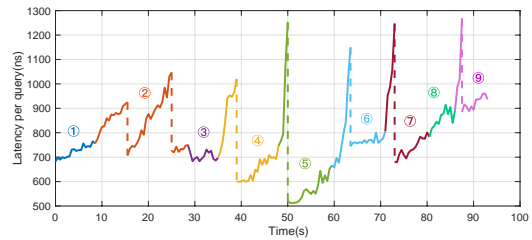

Figure 13: Performance of Incremental Learning

## 5.4 Performance of Incremental Learning

NIS assumes that data and query distributions of an application only change slowly and we propose an incremental learning technique to periodically update our index structure. In this experiment, we evaluate the performance of incremental learning on the uniform dataset with 200 million keys initially. Each experiment is run for 6 episodes, and different workloads are submitted for processing during each episode. The first workload is used to train the NIS model, while the rest 5 are employed to test the performance of incremental learning. A workload contains 5 million lookups and 5 million insertions. So the size of dataset increases from 200 millions to 230 millions with 30 millions newly inserted data. Consequently, the average processing cost of queries will gradually increase. We compare three approaches. The default approach does not apply incremental learning. The *inc* approach shows the performance of NIS using incremental learning, while the *trained* line shows the ideal case that we train a new model from scratch at each episode for all existing data and queries.

In Figure 12(a), we train NIS with a uniform workload, where both reads and insertions are randomly generated based on the data density. All the following 5 workloads are also uniform workloads. This represents the case where query patterns do not change over time, and hence, even default approach shows a performance similar to the ideal case. In Figure 12(b), both reads and writes follow the log-normal distribution (mean=0, stdvar=0.7). We simulate the case where we change data from uniform distribution to log-normal one, while keeping the query distribution as log-normal distribution. We can see that the incremental learning still can provide a similar performance as the ideal one. In Figure 12(c), we simulate the case where both data and query distributions change over time. All writes follow the log-normal distribution, and all reads are randomly sampled from keys based on the data density. During the insertion, the density changes over time, causing the query distribution to evolve as well. We observe that the incremental learning can still tune the index structure, but a bit slower than the ideal case, because it realizes the changes only when a new episode starts. In the last case (Figure 12(d)), we test the case where the query distribution dramatically changes in a short time. The reads follow a normal distribution with a moving center for each episode. Writes follow the log-normal distribution. After a few episodes, the performance of NIS has a large gap from the ideal case. It indicates that a full learning process is required, if data and query distribution change frequently over time.

Finally, we show the progress of incremental learning with a dramatic changing workload in Figure 13. We have an original dataset with 50 million keys and 9 different workloads, each of which contains 10 million lookups and 10 million insertions, and follows a log-normal distribution with a different center. Every 10 seconds, the incremental learning is invoked to update the index. And we can see that the new index structure can provide a better performance until the next workload starts.

## 6 RELATED WORK

The modern database management system (DBMS) becomes so complex for optimization and maintenance, that even database experts may not be able to figure out the optimal design and configuration of the database for specific applications. Recently, the database community starts applying deep learning techniques to reduce the complexity of database management. In [20], some possible research areas for deep learning techniques on database, such as database tuning, query optimization and index search, are discussed.

In the database tuning area, the CMU group designs the OtterTune[2], an autonomous database system [15][25][16]. The OtterTune collects the data for the running status of the database and builds a series of machine learning models (including deep learning models and classic machine learning models) to find the optimal configuration knobs. The idea is to allow anyone to deploy a DBMS without any expertise in database administration. Following their approach, CBDTune[26] proposes to use the reinforcement learning model to perform the configuration tuning for the cloud database. They adopt the deterministic policy gradient model which is similar to the one used in the NIS. The performance change (latency and throughput) is used as the reward during the training process.

Different from OtterTune and CBDTune, Li et. al. propose a query-aware automatic tuning approach, QTune[10]. QTune vectorizes a set of SQL queries by extracting the semantic features of the SQL queries. The vector representations of SQL queries are fed to the deep reinforcement learning model, which is trained to identify the proper configurations of the DBMS optimized for those queries.

Queries involving multiple join operators incur high processing costs and are hard to optimize. Krishnan et. al. propose applying the deep learning techniques for join query optimization[8]. Their reinforcement learning model is integrated with Postgres and SparkSQL and is shown to be able to generate plans with optimization costs and query execution times competitive with the native query optimizer. SkinnerDB[19] further improves the prediction of join queries by splitting queries into many small time slices. The learning model tries different join orders during different time slices and promising plan is selected. Neo[13], on the other hand, tries to rewrite the database optimizer in a learning language. Neo bootstraps its query optimization model from the conventional database optimizers and updates its strategy based on the predicted query plans and their real processing costs. Experiments show that Neo outperforms the original optimizer in Postgres.

---

[2]https://github.com/cmu-db/ottertune

Instead of replacing the whole database optimizer, some work try to improve the cost estimation using the deep learning model. In [22], [24] and [4], deep learning models are applied to estimate the query selectivities or data cardinalities. If columns are highly correlated, the histogram-based estimation may be far from the real result, due to the independent assumption. On the contrary, deep learning model can catch the correlations among columns and rows. Therefore, it can potentially generate a more precise estimation.

The deep learning approach can be also adopted to search for new data structures for the DBMS[6]. The closest work to ours is the learned index from Google[7]. They formalize the index as a key mapping function(given a key, return its position at the disk) and apply the neural models to learn the index in an ad hoc way. We adopt a totally different approach by sticking to the conventional index structures (ordered and unordered blocks in this paper) and ask the neural model to learn how those basic structures can be assembled together as a full-fledged index. Once the index has been predicted, we do not need the neural model. Hence, the performance of index is comparable to state-of-the-art indexes, because we avoid the expensive cost incurred by the neural model prediction in [7].

## 7 CONCLUSION

In this paper, we propose a Neural Index Search(NIS) approach to automatically tune indexes for a given dataset and workload. The NIS applies the reinforcement learning approach to assemble abstract index blocks into a full-fledged index and tests it against the target workload. The index performance is used as the reward for the learning model to update its strategy. Gradually, the predicted index converges to a fine-tuned structure. In theory, our NIS can simulate many existing index structures, such as $B^+$-tree index, Hash index and Skip List index. It can also explore the index structures that have never been examined. We also propose an incremental learning approach to support progressive updates of NIS. In our experiments, the index generated by NIS achieves a comparable performance to existing state-of-the-art index structures.